# Sustained Formation of Progenitor Globular Clusters in a Giant Elliptical Galaxy

Jeremy Lim[1], Emily Wong[1], Youichi Ohyama[2], Tom Broadhurst[3,4,5], & Elinor Medezinski[6]

[1]*Department of Physics, The University of Hong Kong, Pokfulam Road, Hong Kong*

[2]*Academia Sinica, Institute of Astronomy and Astrophysics, 11F of AS/NTU Astronomy-Mathematics Building, No.1, Sec. 4, Roosevelt Rd, Taipei 10617, Taiwan*

[3]*Department of Theoretical Physics, University of Basque Country UPV/EHU, Bilbao, Spain*

[4]*IKERBASQUE, Basque Foundation for Science, Bilbao, Spain*

[5]*Donostia International Physics Center, Paseo Manuel de Lardizabal, 4, San Sebastián, 20018, Spain*

[6]*Department of Astrophysical Sciences, 4 Ivy Lane, Princeton University, Princeton, NJ 08544, USA*

**Globular clusters (GCs) are thought to be ancient relics from the early formative phase of galaxies, although their physical origin remains uncertain[1,2]. GCs are most numerous around massive elliptical galaxies, where they can exhibit a broad colour dispersion, suggesting a wide metallicity spread[3]. Here, we show that many thousands of compact and massive ($\sim 5 \times 10^3$–$3 \times 10^6$ M$_\odot$) star clusters have formed at an approximately steady rate over, at least, the past ~1 Gyr around NGC 1275, the central giant elliptical galaxy of the Perseus cluster. Beyond ~1 Gyr, these star clusters are indistinguishable in broadband optical colours from the more numerous GCs. Their number distribution exhibits a similar dependence with**

**luminosity and mass as the GCs, whereas their spatial distribution resembles a filamentary network of multiphase gas[4,5] associated with cooling of the intracluster gas[6,7]. The sustained formation of these star clusters demonstrates that progenitor GCs can form over cosmic history from cooled intracluster gas, thus contributing to both the large number and broad colour dispersion – owing to an age spread, in addition to a spread in metallicity – of GCs in massive elliptical galaxies. The progenitor GCs have minimal masses well below the maximal masses of Galactic open star clusters, affirming a common formation mechanism for star clusters over all mass scales[8–10] irrespective of their formative pathways.**

Images taken with the Advanced Camera for Surveys (ACS) on the NASA/ESA Hubble Space Telescope (HST) are able to separate, and even resolve, individual star clusters in NGC 1275. We have identified all the star clusters detectable towards this galaxy (see Methods) from images taken in three filters: F435W in the blue ($B$), F550M in the middle of the visual band ($V$), and F625W in the red ($R$). Previous studies have been restricted to just the innermost regions[11–13] of NGC 1275, or employed the same images as we do but focused mainly on the outskirts[14,15] (see summary in Supplementary Information). Although finding numerous young, compact, and massive star clusters interpreted as progenitor GCs, these studies did not establish their full properties (total number, overall spatial distribution, full age spread, luminosity or mass function, and formation rate over time) nor their relationship with other galaxies along the sightline towards NGC 1275. In our work, we identify star clusters throughout the $3'.4 \times 4'.0$ (73.4 kpc × 86.4 kpc) field of view common to all three filters. In this way, we are able to separate out star clusters belonging to other galaxies, and determine in full the properties of those definitively belonging to

NGC 1275.

Fig. 1 shows a colour-colour diagram of the star clusters detected in all three filters excluding (see Supplementary Fig. 1 including) those bounded by two galaxies along the sightline towards NGC 1275 (see Supplementary Fig. 2 for their number distribution with colour, and Supplementary Table 1 for object numbers). Their ages can be inferred from theoretical evolutionary tracks for coeval stellar populations provided their metallicity (elemental abundance not in hydrogen and helium) is known. Fig. 1$b$ shows evolutionary tracks (see Methods) for star clusters having metallicities of the intracluster gas, 0.4 $Z_\odot$ [ref. 16], and the Solar metallicity, 1.0 $Z_\odot$. Star clusters become redder with age as progressively less massive stars evolve away from the main sequence, resulting in an increasing ($B-V$) and ($V-R$) with age. Those having ages ≲6 Myr are bluest in ($B-V$) but relatively red in ($V-R$), however, owing to a H$\alpha$+[NII] emission from leftover gas photoionized by massive stars (i.e., H II regions) that contributes to the $R$ filter. Such massive stars have short lifetimes on the main sequence (~1–10s Myr, depending on their mass and metallicity), after which they evolve to become red supergiants before being destroyed in supernova explosions to create the loop in the evolutionary tracks between ~7 Myr and ~15 Myr for 0.4 $Z_\odot$ and between ~7 Myr and ~50 Myr for 1.0 $Z_\odot$. A striking feature of this colour-colour diagram is the very broad and continuous range in ($B-V$) spanned by the star clusters, covering the full range of ages spanned by the stellar evolutionary tracks from the present epoch back to the early Universe.

Before assigning ages to individual star clusters, we first have to separate out GCs, which can have a metallicity different from the values adopted for the evolutionary tracks in Fig. 1. GCs generally exhibit a bimodal colour distribution, indicating distinct metal-poor (bluer) and metal-rich (redder) populations that differ by about a factor of ten in metallicity[1, 2]. In giant elliptical galaxies, however, the colour distribution of GCs can be continuous rather than bimodal[3] (e.g., Supplementary Figs.9 and 10); if GCs constitute a single ancient population ($\gtrsim$10 Gyr old), then their broad colour dispersion translates to metallicities spanning the enormous range $\sim$0.005 $Z_\odot$ 1.0 $Z_\odot$ [ref. 3]. Even then, GCs are commonly divided into two populations having different average metallicities. In Fig. 1$b$, the colour of 10-Gyr star clusters having a metallicity of 0.02 $Z_\odot$ representative of metal-poor GCs is indicated by a blue filled circle, whereas those having a metallicity 0.4 $Z_\odot$ representative of metal-rich GCs is indicated by a red filled circle (see Methods for computations of colours). There is a clear concentration of star clusters in these regions of the colour-colour diagram[12, 14], as would be expected for GCs belonging to NGC 1275.

The star clusters in Fig. 1 can be separated into two groups in colour-colour space depending on whether their spatial distribution is isotropic (except towards other galaxies along the same sightline as NGC 1275) or anisotropic. Fig. 2$a$ shows the locations of all those cataloged lying in the region of the color-color diagram coded blue, whereas Fig. 2$b$ shows the locations of all those cataloged lying in the region of the color-color diagram coded red, in Fig. 1. The green contour in Fig. 2 traces the outermost H$\alpha$ extent of a foreground spiral galaxy known as the High Velocity System (HVS), which is moving almost exactly along our sightline towards NGC 1275 [ref. 17]. There is an appreciable concentration of star clusters, coded both blue and red, along the dusty spiral arms of the HVS (see closeup in Supplementary Fig. 5), indicating that at least some the star clusters towards the HVS actually belong to this galaxy. Excluding the region bounded by the HVS (green contour), the star clusters coded red (Fig. 2*b*) have an isotropic distribution that decreases radially outwards in surface density (Supplementary Fig. 8), as is characteristic of GCs[1, 2]. On the other hand, the star clusters coded blue (Fig. 2*a*), hereafter referred to as blue star clusters (BSCs), have a complex spatial distribution. Away from the central region of NGC 1275, they are spatially concentrated in patterns related to the morphology of the gaseous nebula in this galaxy, and must therefore belong to NGC 1275.

Our ability to separate the star clusters into two distinct groups depends on whether they are cleanly separated in colour, and also on photometric measurement uncertainties. As can be seen from Fig. 2, we deliberately separated the two such that no spatial anisotropy is appreciable among the GCs (except towards the HVS). As a consequence, there is almost certainly a small degree of contamination by GCs among the BSCs, as evidenced by the low surface density of isotropically distributed star clusters in the outer regions of Fig. 2a. Any contamination by BSCs among the GCs, however, would be lost against the far more numerous GCs. To mitigate against contamination by GCs among the BSCs, all the star clusters lying beyond the grey contour in Fig. 2a have been excluded from Fig. 1. Furthermore, to avoid contamination by star clusters belonging to the HVS, all those (both BSCs and GCs) in the region bounded by the green contour have been omitted from Fig. 1. Those in the central region bounded by the cyan contour in Fig. 2 (see closeup in Supplementary Fig. 6) have likewise been omitted, as they likely belong to yet another spiral galaxy along the sightline towards NGC 1275 (see Supplementary Information).

Fig. 3 shows the luminosity or, by translating from luminosity to mass based on the stel-lar evolutionary model employed (see Methods), mass function (assuming 0.4 $Z_\odot$) of the BSCs from Fig. 1 over the age range ∼100–900 Myr (see Supplementary Fig. 7). These BSCs are separated into three age brackets having nearly equal $(B-V)$ colour widths (Supplementary Table 4) of ∼100–400 Myr, ∼400–600 Myr, and ∼600–900 Myr (see Supplementary Table 3 for cross contamination among the different age bins); the number of BSCs in the middle age bracket has been multiplied by a factor of 1.5 to facilitate a direct visual comparison with the other age brackets. Also shown are the luminosity and mass functions of the GCs in Fig. 1, divided into metal-poor and metal-rich populations at $(B-V) = 0.917$ (see Methods). The luminosity functions of GCs can be universally fit by log-normal functions[1,2,18] with a peak just above ∼$10^5 L_\odot$, which is close to the detection threshold where our source identification becomes incomplete. To permit an instructive comparison between the BSCs and GCs, we fit simple power laws (straight lines in Fig. 3) to the luminosity and mass functions above their respective detection thresholds; i.e., $\partial N/\partial L \propto L^{-\alpha}$ or $\partial N/\partial M \propto M^{-\beta}$, $\partial N$ being the number of star clusters over the luminosity interval $L$ to $L + \partial L$ or mass interval $M$ to $M + \partial M$. Both the BSCs and GCs exhibit a similar functional dependence of $\alpha \simeq 1.9$–2.1 and $\beta \simeq 2.0$–2.2 (Supplementary Table 2). Furthermore, in a past study[19] of over two-hundred BSCs in NGC 1275, their radial light profiles have been found to resemble GCs. At a given luminosity or mass interval, the BSCs over the age range ∼100-900 Myr are approximately one-tenth or one-hundredth, respectively, as numerous as the GCs over the sky area searched.

The GCs in NGC 1275 have maximal masses of ∼$10^7 M_\odot$, similar to the most massive GCs found in other giant elliptical galaxies[18]. They also are similar in other ways, exhibiting a mass-metallicity relationship[1,2] and a more centrally concentrated metal-rich than metal-poor population[1, 2] (Supplementary Figs. 9–11). The BSCs, by comparison, have maximal masses of $\sim 10^6\,M_\odot$, about an order of magnitude higher than the typical GC mass at the peak in their number distribution[1,2], and about two orders of magnitude higher than the maximal masses of open star clusters in our Galaxy. Their mass function is approximately constant across the three age intervals plotted in Fig. 3, indicating an approximately steady formation rate over the past $\sim 1$ Gyr. Over the mass range $\sim 5\times 10^4 - 1\times 10^6\,M_\odot$ for the BSCs spanning ages $\sim 100$-$900$ Myr, their formation rate is $1.2\,\mathrm{Myr}^{-1}$ at a total mass rate of $1\times 10^5\,M_\odot\,\mathrm{Myr}^{-1}$; i.e., one star cluster born every 1 million years having a mass close to that of GCs at the peak in their number distribution. These values are lower limits as some bounded by the HVS likely belong to NGC 1275, and furthermore we do not include the even more numerous BSCs detected in only one or two filters and likely having lower individual masses (Supplementary Figs 3 and 4, and Supplementary Table 1). We cannot rule out temporal modulations in the formation rate of BSCs on timescales shorter than a few hundred Myr; the mass formation rate of the youngest BSCs ($<10^8$ yr) has been estimated to be 1–2 orders of magnitude higher[14,15]. Although these star clusters have lower bounds for their masses that are an order of magnitude smaller ($\sim 5\times 10^3\,M_\odot$), this difference alone can only account (given the same mass function) for a factor of less than 2 difference in the total mass formation rate.

The close spatial relationship between the BSCs and the nebula in NGC 1275 indicates that these star clusters formed from the nebular gas[14, 15], which has an exact counterpart in molecular gas[4,5] – the reservoir for star formation. Such nebulae are commonly found in the central galaxies of virialized clusters[6,7], where the intracluster gas (at temperatures $\sim 10^{7\text{-}8}$ K) at the cluster core is especially dense and experiences intensive radiative cooling.

The spatial distribution of the BSCs, at birth related to the morphology of the nebula, will become more isotropic over time as the BSCs orbit the galaxy. Lower-mass BSCs are predicted to more quickly dissolve owing to stellar mass-loss and two-body relaxation in the tidal field of their host galaxy[20] (Supplementary Information), the mechanism held responsible for the low-mass turnover among ancient GCs[1–3]. Thus, over time, both the spatial distribution and the luminosity and hence mass distributions of the BSCs will increasingly resemble the log-normal distribution of GCs. Well before that (see Supplementary Information for survival timescales), the BSCs will already have evolved to become observationally indistinguishable in broadband optical colours from the metal-poor GCs, by an age of just $\sim$1–2 Gyr, and thereafter the metal-rich GCs, by an age of $\gtrsim$3 Gyr. Fading by just 1 mag between 1 Gyr and 3 Gyr, the BSCs will then constitute, by luminosity, $\sim$5–10% of the GC population, thus contributing to the colour dispersion of presumed ancient GCs.

The progenitor GCs in NGC 1275 and those in nearby spiral and irregular galaxies[21–24] experiencing a brief period of highly elevated star formation (starbursts) due to mergers share a similar dependence in their luminosity[9, 22, 24] and, where measured, mass functions[9, 25], as do lower-mass open star clusters in our Galaxy[26] and other spiral galaxies[27, 28]. This universality suggests a common formation mechanism for star clusters over all mass scales[8–10], as well as through entirely different formative pathways. Unlike in merging galaxies, in cluster central galaxies the gas reservoir for star formation can be replenished from cooled intracluster gas so as to sustain the formation of progenitor GCs over cosmic history.

**Acknowledgements** Based on observations made with the NASA/ESA Hubble Space Telescope, and obtained from the Hubble Legacy Archive, which is a collaboration between the Space Telescope Science Institute (STScI/NASA), the Space Telescope European Coordinating Facility (ST-ECF/ESA) and the Canadian Astronomy Data Centre (CADC/NRC/CSA). J. L. acknowledges support from the Research Grants Council of Hong Kong through grants 17303414 and 17304817. Y. O. acknowledges support by grant MOST 107-2119-M-001-026-. T. B. thanks HKU for generous support from the Visiting Research Professor Scheme.

**Contributions** J. L. supervised the project, and wrote the paper. E. W. and Y. O. conducted the technical aspects of the work. T. B. and E. M. initiated the project, and participated in the interpretation of the results.

**Competing interests** The authors declare that they have no competing financial interests.

**Corresponding author** Correspondence and requests for materials should be addressed to J.L. (email: jjlim@hku.hk).

**Data availability statement** The data that support the plots within this paper and other findings of this study are available from the corresponding author upon reasonable request.

**Methods**

**Data.** We obtained fully processed, science ready, images of NGC 1275 taken through the F435W (*B*), F550M (*V*), and F625W (*R*) filters with the ACS on the HST from the Hubble Legacy Archive (proposal id. 10546). Exposure times were 9834 s in *B*, 9714 s in *V*, and 9924s in *R*. Zero-points for converting between instrumental and Vega magnitudes in these HST filters can be found in the ACS data handbook (http://www.stsci.edu/hst/acs/documents/handbooks/currentDHB/acs_cover.html). The *B*, *V*, and *R* designations that we use therefore refer to the relevant HST filters, and not to the standard Johnson-Cousins filters in *B*, *V*, and *R*.

**Compilation and Photometry of Candidate Star Clusters.** The complex light distribution at the inner region of NGC 1275 – exhibiting complicated dust features (Fig. 2*b* and Supplementary Fig. 5) and a central spiral-like morphology (Supplementary Fig. 6) – creates challenges in compiling a complete list of star clusters in this galaxy. Previous studies have found the BSCs and GCs in NGC 1275 to be only poorly resolved[12,19], as would be expected for a typical open star cluster or GC at the distance to this galaxy. Given the close angular separation between star clusters as is evident in Fig. 2, the situation is therefore closely analogous to doing photometry in crowded star fields (over a varying background).

We used StarFinder[29] (http://www.bo.astro.it/StarFinder/), an algorithm for crowded stellar fields analysis, to find and conduct photometry of compact objects – candidate star clusters – towards NGC 1275. This algorithm requires the point-spread-function (PSF) of an image, which we obtained by averaging the images of several bright stars taken through a particular filter. Pho-tometry in each filter comprises PSF fitting to an object, along with a fit for the local background as represented by a slanting plane. In the $R$ filter where the nebular gas also contributes light, this approach helps remove any nebular light from the object photometry. For each object found, StarFinder returns a correlation index ranging from 0 to 1, with larger values corresponding to a closer match with the PSF. The number of objects found peak at a correlation index of 0.97 in $B$ and $V$ and 0.98 in $R$, and decrease rapidly with decreasing correlation index before levelling out at a value of ∼0.5. We therefore imposed a correlation index of >0.5 for objects compiled in our master catalog. Furthermore, we set a local detection threshold of $4\sigma$ for each object, where $\sigma$ is the local root-mean-square noise level.

**Exclusions of Contaminants.** We visually inspected the objects found, especially those in the bright inner region and near bright stars, galaxies, and the image boundaries, to both gauge the performance of the algorithm and to exclude from our master catalog objects that are clearly or likely artefacts. Supplementary Fig. 1 shows a colour-colour and a colour-magnitude diagram of all the objects compiled in our master catalog that were detected in all three filters (each dot representing a single object), numbering a total of 12944 objects. The brown dots, isolated from the others in both the colour-colour and colour-magnitude diagrams, have the colours of M stars and hence comprise Galactic M stars. The magenta dots are not as well isolated in the colour-colour diagram, but clearly isolated in the colour-magnitude diagram as relatively bright objects compared with the majority having the same colours. Those having colours $(B-V) < 1$ are primarily located at the north-western part of the central spiral galaxy, which lie behind dust associated with the HVS (see Supplementary Fig. 6a); these objects are presumably reddened star clusters associated with the central spiral galaxy. Those having colours $(B - V) > 1$ are isotropically distributed except for a few lying at the north-western part of the central spiral galaxy: the colours of these objects are intermediate between the reddest GCs and the brown dots (Galactic M stars), and away from the central spiral correspond primarily to Galactic M stars. All these objects were excluded in our final catalog, hereafter referred to as the source catalog, which contains a total of 12729 objects detected in all three filters (see Supplementary Table 1). Although we cannot completely rule out contamination by dim Galactic stars or dim and compact galaxies, any such contamination must be small as evidenced by the highly structured spatial distribution of the BSCs (Fig. 2a) and the outwardly decreasing radial surface density of the GCs (Supplementary Fig. 7).

**Correction for Galactic Extinction.** We applied the same corrections for Galactic extinction as described in ref. 13, amounting to 0.67 mag in $B$, 0.50 mag in $V$, and 0.43 mag in $R$, using the reddening law of ref. 30.

**Theoretical Evolutionary Tracks.** To compute how the colours of a star cluster change with age, as well as the colours of ancient GCs, we used the Yggdrasil algorithm[31] (http://ttt.astro.su.se/projects/yggdrasil/yggdrasil.html) for single (i.e., coeval) stellar populations as taken from the Starburst99 software and data package[32] (http://www.stsci.edu/science/starburst99/docs/default.htm). We assumed a Kroupa[33] initial mass function (IMF) truncated beyond the stellar mass range 0.1–100$M_\odot$. Massive stars ($>10M_\odot$) dominate the light from young star clusters but have short lifetimes, and do not affect the evolutionary track beyond an age of a few tens of Myr. When present, however, they can ionise gas leftover from star formation to create a H II region. We assumed a unity filling factor for such leftover gas, which primarily affects the brightness in $R$ through the H$\alpha$+[N II] lines, so as to reproduce the broad range in $(V-R)$ exhibited by the youngest BSCs. (Other bright lines from H II regions such as the H$\beta$ and [O III] $\lambda\lambda$495.9, 500.7 nm lines fall outside the HST filters utilised in our study). The evolutionary tracks in the colour-colour diagram of Fig. 1 have been taken through the responses of the HST F435W, F550M, and F625W filters (see http://www.stsci.edu/hst/acs/documents/handbooks/currentDHB/acs cover.html), and also take into consideration the redshift of NGC 1275.

**Luminosity to Mass Conversion.** We used the Yggdrasil algorithm, which provides the rest-frame spectrum (in absolute magnitude) of a star cluster for a given age, metallicity, IMF, and a fiducial mass of $10^6 M_\odot$. For the GCs, we adopted a metallicity of $Z = 0.02 Z_\odot$ for the metal-poor and $Z = 0.4 Z_\odot$ for the metal-rich populations. Furthermore, we assumed a uniform age of 10 Gyr, and a Kroupa IMF truncated beyond the stellar mass range 0.1–100$M_\odot$. The GCs are separated into metal-poor and metal-rich populations by simply imposing a dividing line at $(B-V) = 0.917$, midway between their predicted colours for the two selected metallicities. For the BSCs, we inferred their ages by comparing their measured nominal $(B-V)$ values with those predicted by the Yggdrasil algorithm assuming $Z = 0.4 Z_\odot$ and a similarly truncated Kroupa IMF. Applying a Doppler shift to the theoretical spectrum given a redshift of $z = 0.0176$ for NGC 1275, we passed the theoretical spectrum of a given star cluster through the passband of a particular filter to compute the predicted absolute magnitude of that cluster in that filter. Assuming a distance to NGC 1275 of 74 Mpc, conversion between luminosity and mass requires a simple scaling between the observed and predicted absolute magnitudes to the fiducial mass of that cluster.

**Data availability statement** The data that support the plots within this paper and other findings of this study are available from the corresponding author upon reasonable request.

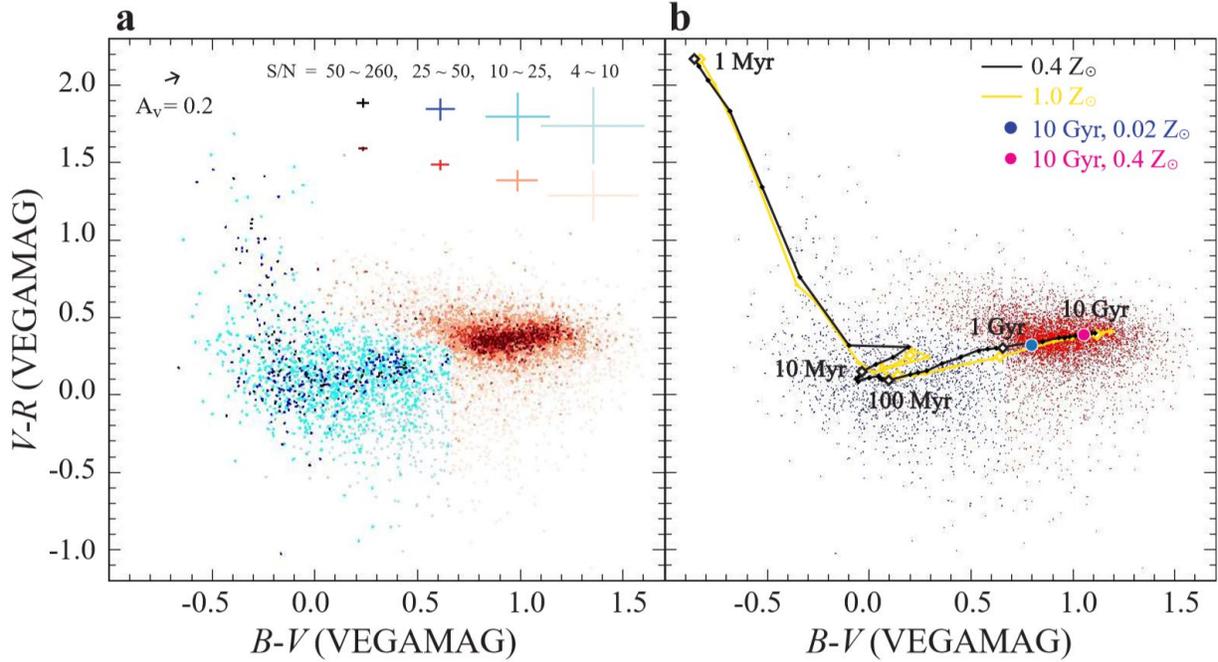

Figure 1: **Colour-colour diagram of star clusters belonging to NGC 1275.** Each dot represents an individual star cluster detected in all three HST filters F435W ($B$), F550M ($V$), and F625W ($R$) (see Methods for conversion to Vega magnitude). **a**, Different shades have different average colour measurement uncertainties, as denoted by crosses with arm lengths of $\pm 1\sigma$, corresponding to the different ranges in signal-to-noise ratios (S/N) indicated. Blue dots are relatively young star clusters referred to as blue star clusters (BSCs), and red dots are globular clusters (GCs); these two populations have different global spatial distributions (see Fig. 2). The vector indicates an extinction of $A_V$=0.2, corresponding to the average dust extinction required to systematically displace the BSCs having ages ≲6 Myr (which exhibit HII regions) from the theoretical evolutionary tracks. There is no appreciable dust extinction towards the other BSCs plotted in this figure (see Supplementary Information).   **b**, Theoretical evolutionary tracks for star clusters having metallicities of $0.4\,Z_\odot$ in black and $1.0\,Z_\odot$ in yellow, with equally spaced time intervals marked on each track for every decade in age (see Methods). The solid blue and red filled circles indicate the colours of 10-Gyr star clusters having metallicities of $0.02\,Z_\odot$ and $0.4\,Z_\odot$, respectively, corresponding to the approximate average metallicities of metal-poor and metal-rich globular clusters, respectively, in giant elliptical galaxies[3]. The star clusters with the highest photometric accuracies (smallest colour uncertainties) are tightly grouped along the evolutionary tracks. Star clusters in regions bounded by intervening galaxies as indicated by the green and cyan contours in Fig. 2, as well as those lying in the same region of the colour-colour diagram as the blue dots but beyond the grey contour in Fig. 2a, have been omitted from Fig. 1 (see reasons in text).

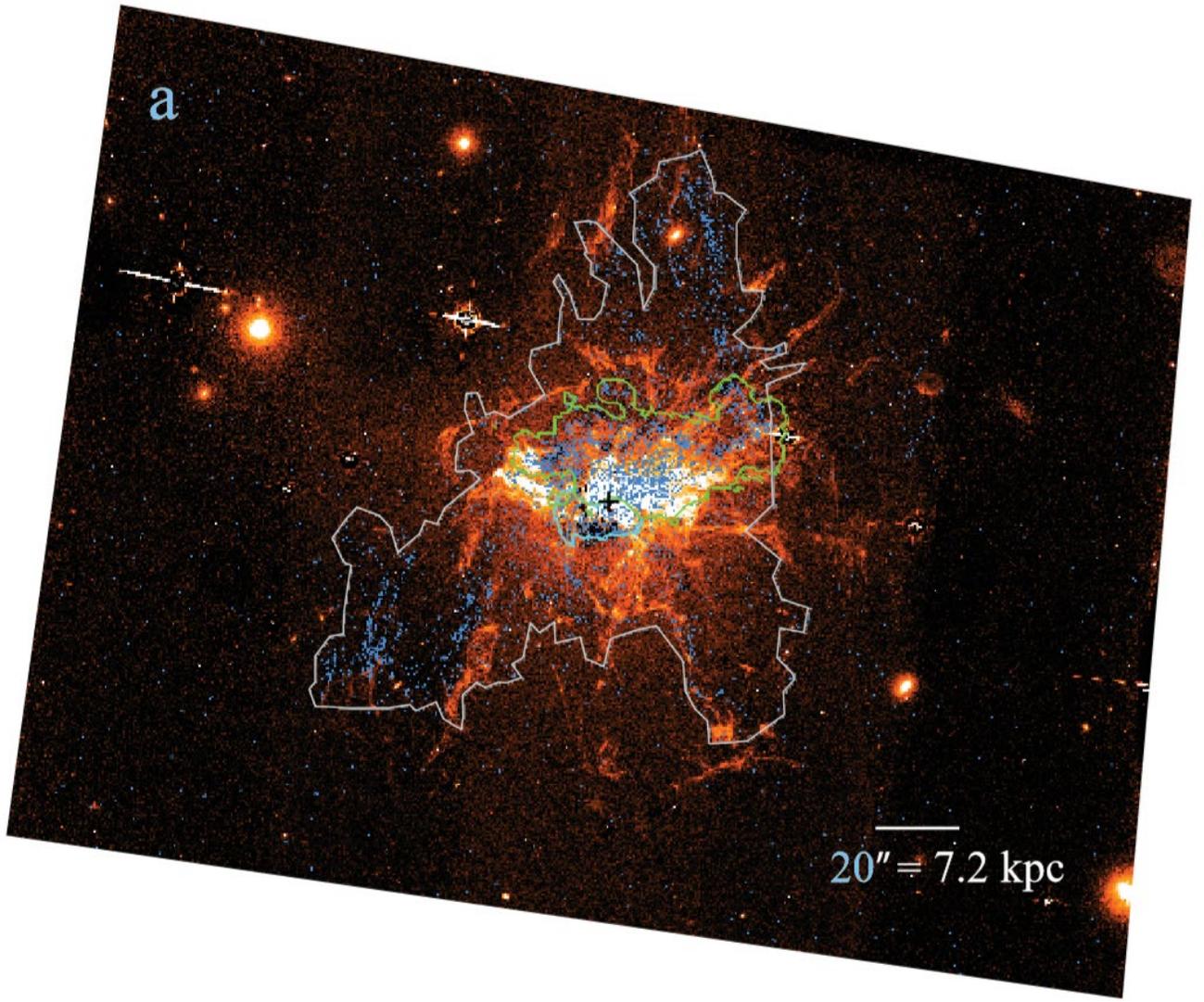

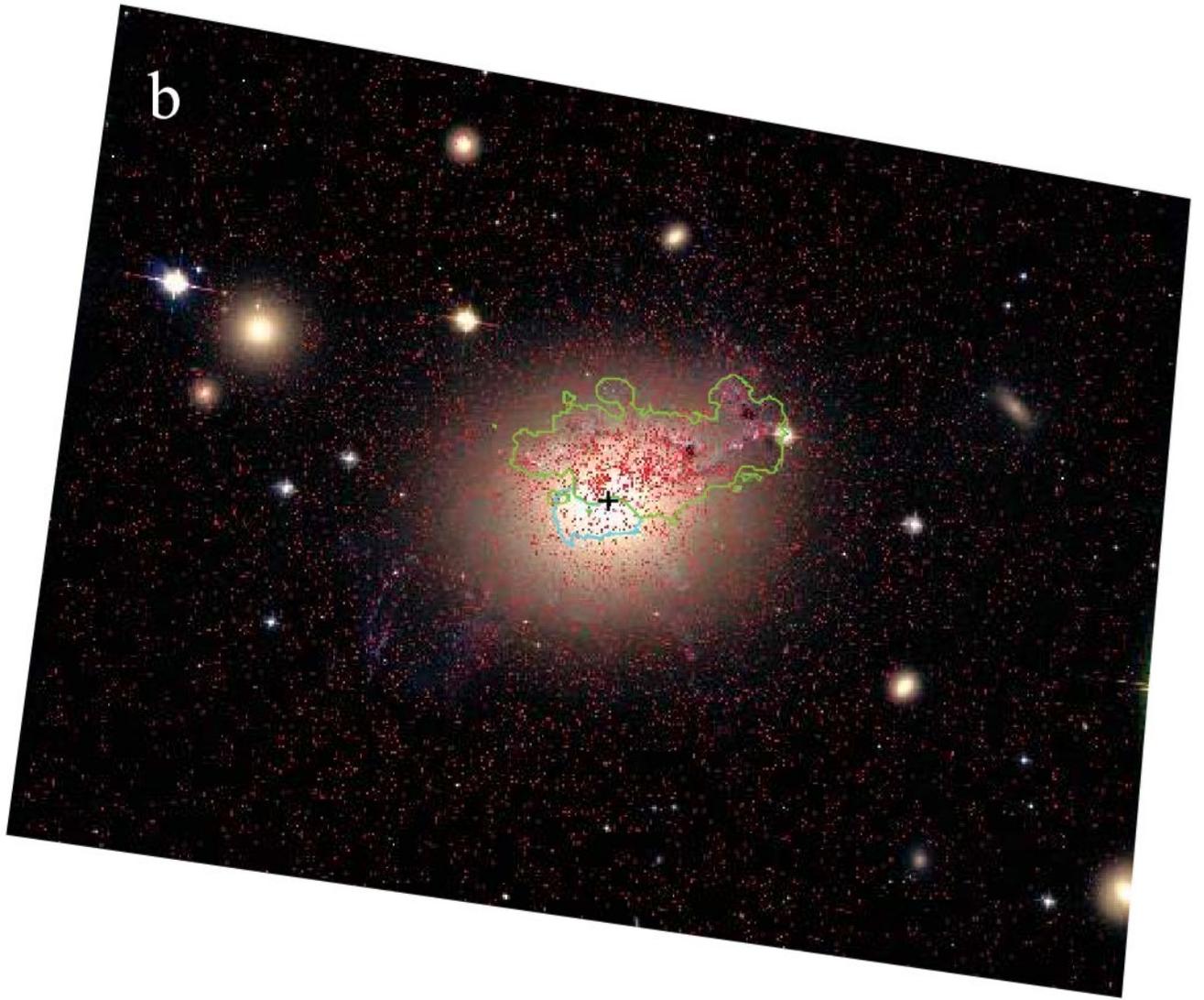

Figure 2: **Spatial distributions of different star cluster populations in NGC 1275.** Each dot indicates the spatial location of a star cluster lying in the region of the colour-colour diagram shown in Fig. 1 coded **a**, blue (referred to as blue star clusters, or BSCs) overlaid on a H$\alpha$+[NII] image of the gaseous nebula in NGC 1275; **b**, red (globular clusters, or GCs) overlaid on a multi-colour (*BVR*) image of the same region. The green contour traces the outermost extent of a foreground infalling galaxy[17] known as the High Velocity System (HVS); the cyan contour traces the outermost visible extent of spiral arms related to yet another galaxy (see Supplementary Information). The cross indicates the nucleus of NGC 1275. The BSCs in panel **a** are strongly concentrated in distinct spatial patterns, by contrast with the GCs in panel **b** that, away from the HVS, have an isotropic distribution. BSCs lying beyond the grey contour in panel **a** appear to have a more isotropic distribution, and likely suffer to some degree from contamination by GCs. GCs lying within the green contour are concentrated towards the dusty arms of the HVS, and a fraction likely comprise extincted and hence reddened young star clusters rather than GCs. To mitigate against contamination by star clusters belonging to other galaxies and cross contamination, all the star clusters lying within the green and cyan contours, and BSCs lying beyond the grey contour, have been excluded from Figs. 1 and 3.

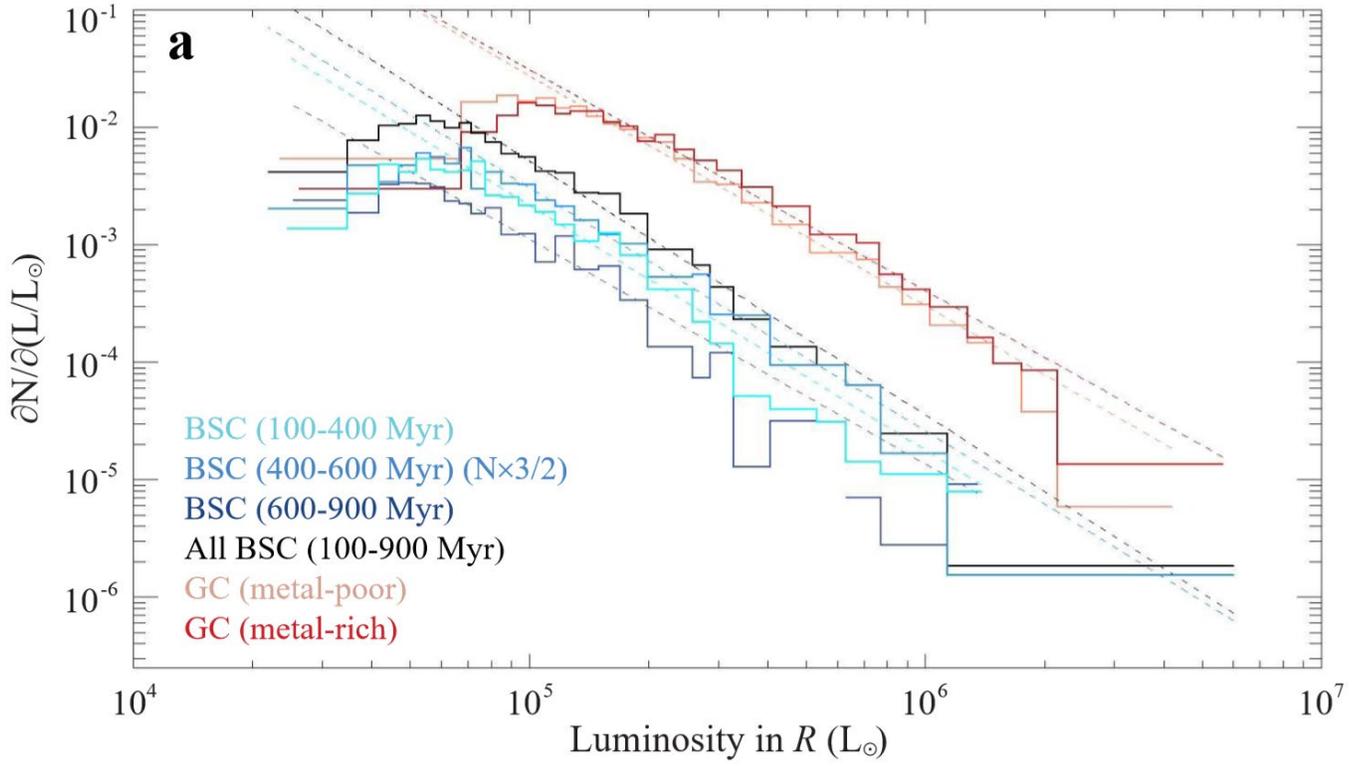

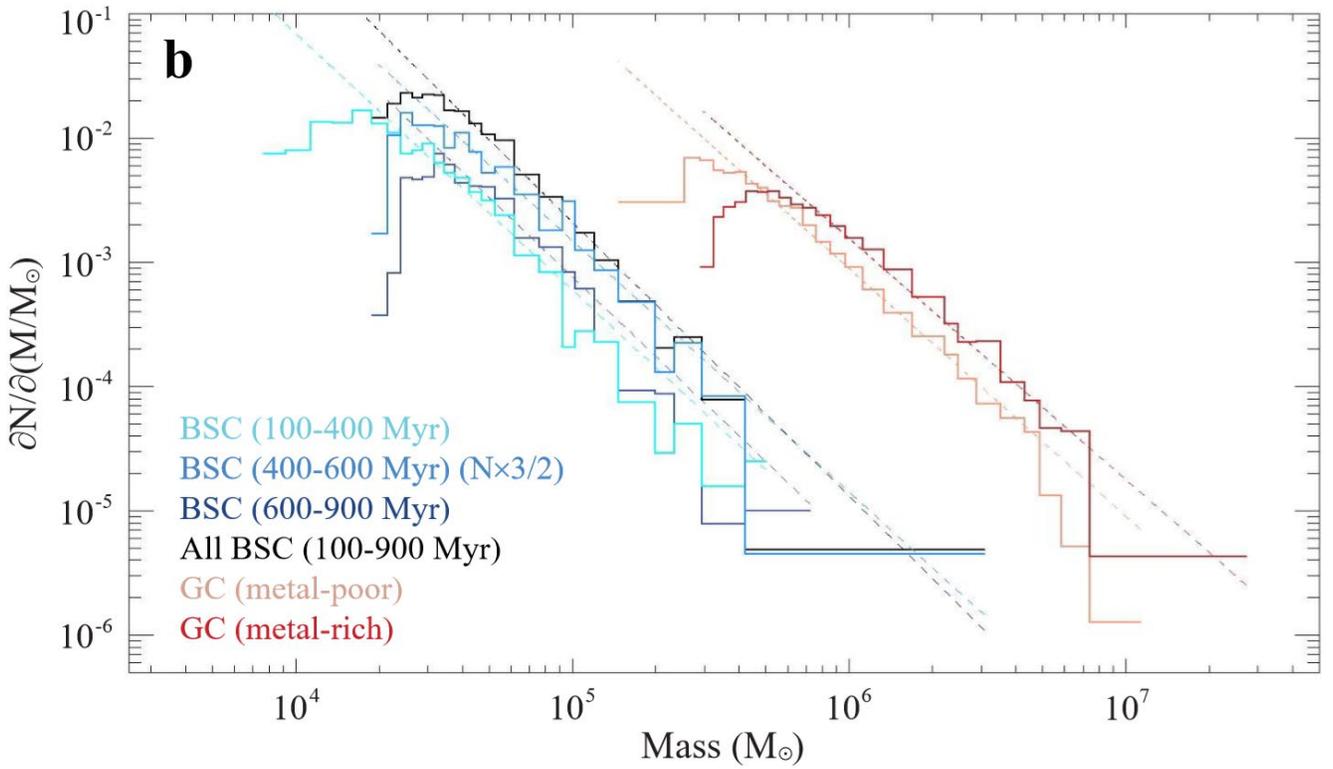

Figure 3: **Luminosity and mass functions for the BSCs and GCs.** Number of star clusters, $N$, in a given: **a**, luminosity interval $L$ to $L+\partial L$ (in units of $L_\odot$, the solar luminosity) as measured in $R$ (F625W); and **b**, mass interval $M$ to $M+\partial M$ (in units of $M_\odot$, the solar mass). These functions are plotted in different shades of blue for the BSCs spanning, for a metallicity of 0.4 $Z_\odot$, the age range ∼100–900 Myr ($0.095 \leq (B-V) \leq 0.621$); and in different shades of red for the metal-poor and metal-rich GCs separated at $(B-V) = 0.917$, midway between the two filled circles plotted in Fig. 1. For BSCs spanning the age range 400-600 Myr, both the luminosity and mass functions have been shifted vertically upwards by a factor of 1.5 so as to provide a direct comparison with the corresponding functions in the two other broader age brackets plotted. The turnover at low luminosities and therefore masses reflect incompleteness near the detection threshold, which varies with spatial location in NGC 1275. The dashed lines are best-fit power-laws to the individual luminosity or mass functions well above the detection threshold, with the power-law indices summarised in Supplementary Table 2.